# Assessing the Performance and Impact of PV Technologies on Storage in Hybrid Renewable Systems

Sharaf K. Magableh, Oraib Dawaghreh, Xuesong Wang, Caisheng Wang
*Department of Electrical and Computer Engineering, Wayne State University, Detroit, United States*
sharaf.magableh@wayne.edu, oraib.dawaghreh@wayne.edu, xswang@wayne.edu, cwang@wayne.edu

*Abstract*—Traditional monofacial photovoltaic (mPV) systems are commonly adopted and well-documented because of their lower upfront costs in comparison to bifacial photovoltaic (bPV) systems. This study investigates how PV technologies impact energy storage in grid-scale hybrid renewable systems, focusing on optimizing and assessing the performance of mPV and bPV technologies integrated with pumped storage hydropower. Using Ludington City, Michigan as a case study and analyzing real-world data such as solar irradiance, ambient temperature, and utility-scale load profiles, the research highlights the operational and economic benefits of bPV systems. The results reveal that bPV systems can pump approximately 10.38% more water annually to the upper reservoir while achieving a lower levelized cost of energy ($0.0578/kWh for bPV vs. $0.0672/kWh for mPV). This study underscores the outstanding potential of bPV systems in enhancing energy storage and management strategies, contributing to a more sustainable and resilient renewable energy future.

## I. Introduction

The push towards achieving zero emissions by 2050 has greatly sped up the expansion of renewable energy and technologies, particularly solar power as the fastest-growing energy source. In the last ten years, the solar industry has made strides surpassing advancements seen in the previous four decades in recognition of technological innovations, and cost reductions [1]. Among the various types of PV technologies, conventional monofacial PV (mPV) panels have limited efficiency because they only capture sunlight from the front side, which means they cannot utilize all the available solar irradiance. As a result, their ability to generate electricity is reduced [2]. Bifacial PV (bPV) panels, by contrast, capture sunlight from both sides, enhancing energy output. This dual-sided absorption also benefits from reflected radiation, further boosting electricity production. Hence, both industry and academia have increasingly turned their attention toward bPV technology, which is projected to see a global market share growth from 30% in 2022 to around 70% by 2030 [3]. Given these advancements, a critical question arises: what will be the impact of these evolving PV technologies on energy storage systems, such as pumped storage hydropower (PSH)?

Limited number of studies have examined the effects of different PV technologies on system costs, environmental impact, and reliability. In [4], the authors provided a detailed review of various energy storage systems that can be integrated with PV, including both electrical and thermal storage options. They also discussed PV-energy storage integration in smart buildings and highlighted energy storage's role in PV system development. However, they did not examine how different types of PV systems affect these energy storage systems or compare their impacts. In [5], the researchers presented a comprehensive comparison between on-grid mPV and bPV systems, evaluating aspects such as energy production, reliability, economic feasibility, ecological impact, and footprint area. Their findings showed that the loss of power supply probability for mPV systems was slightly higher at 0.634% with 12,253 PV panels, compared to bPV systems, which had a probability of 0.576% with 10,924 panels. The total number of panels needed was 12,940 for mPV and 11,584 for bPV. The results indicated that bPV systems generate more power with fewer panels, requiring a smaller installation area, and thus achieving a lower levelized cost of energy (LCOE) over the project's lifetime compared to mPV systems, with LCOE values of around 0.272 for mPV and 0.246 for bPV. However, the study did not consider the integration of any energy storage systems, analyze the impact of different PV technologies on these storage systems, or compare results with other solar technologies and energy storage systems. The impact of different PV modules was thoroughly investigated in [6]. The authors investigated different PV modules, focusing on a hybrid solar array and PSH system. This inspires the current study to include various PV technologies and assess their impacts on system sizing and energy storage optimization.

Based on the literature and the identified research gaps, this study aims to advance understanding in the field of hybrid solar energy systems by offering several key contributions. The core idea of this work is to provide a comprehensive comparison of grid-connected hybrid solar energy systems, specifically focusing on mPV and bPV technologies integrated with PSH systems. The primary goal is to evaluate the influence and analyze how these PV technologies impact the sizing and energy management of energy storage systems with the Ludington pumped storage plant as an example. The simplified diagram of the system under investigation is shown in Fig. 1, including both mPV and bPV technologies combined with the Ludington PSH.

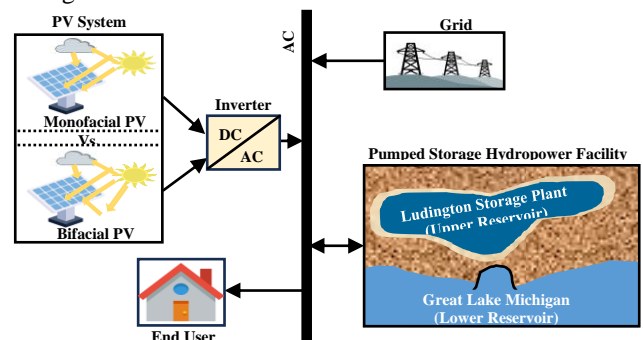

Fig. 1. The schematic diagram of the system under study with different PV technologies combined with a PSH on-grid system for the chosen location.

## II. System Data Acquisition and Processing

Fig. 2. illustrates the variation in residential electricity demand, measured hourly, for a 22kV sub-feeder in Ludington collected from [7]. It shows both the complete load profile for the year 2022 as well as a sample of the detailed view of consumption patterns during the first week of March, highlighting the residential daily load fluctuations. This data is used to design and analyze various PV technologies for grid systems as a case study. Fig. 2 presents these values for the entire year, while Table I highlights MW's minimum, maximum, and average load demand values.

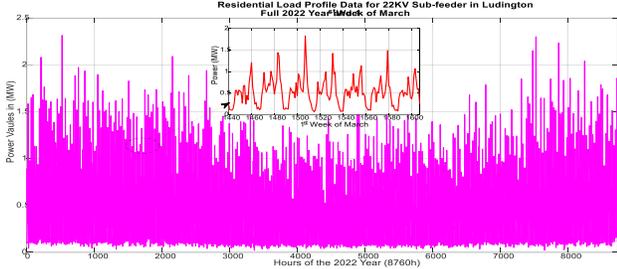

Fig. 2. Hourly residential load profile data for a 22kV sub-feeder in Ludington, Michigan for the full year of 2022 and the first week of March.

For the PV system modeling, data on global horizontal irradiance (GHI), direct normal irradiance (DNI), and diffuse horizontal irradiance (DHI) in $W/m^2$, as well as solar ambient temperature ($T_{amb}$) in °C, were sourced from the National Solar Radiation Database (NSRDB) [8]. These measurements were specifically taken for the Ludington PSH plant Territory in Michigan, USA, to effectively simulate the proposed methodology illustrated in Fig. 3. For detailed analysis, Table I present the maximum, mean, and minimum values for GHI, DNI, and DHI, along with $T_{amb}$ for the year 2022 at this location.

TABLE I. Minimum, average, and maximum recorded data values for the selected location in 2022.

| Acquired Measured Data | | Value | | |
|---|---|---|---|---|
| | | Min | Mean | Max |
| Solar irradiances | GHI ($W/m^2$) | 0 | 153.0991 | 1,002 |
| | DNI ($W/m^2$) | 0 | 163.6844 | 998 |
| | DHI ($W/m^2$) | 0 | 58.8176 | 490 |
| $T_{amb}$ (°C) | | -14.5 | 8.2508 | 27.8 |
| Residential Load Profile (MW) | | 0.0217 | 0.4686 | 2.3153 |

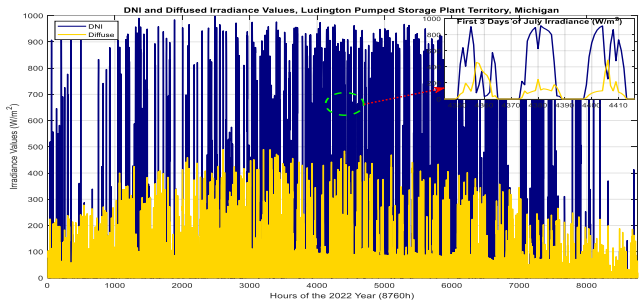

Fig. 3. Annual input irradiance profiles (DNI, and Diffused) for Ludington PSH.

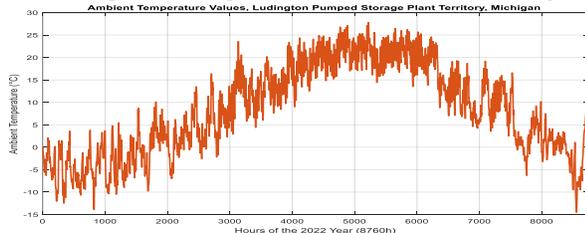

Fig. 4. Annual ambient temperature data for Ludington PSH Territory in 2022.

Fig. 4 shows the ambient temperature in °C of the chosen case study. Despite Michigan's generally colder climate, with ambient temperatures ($T_{amb}$) reaching a maximum of 27.8°C in 2022, this study aims also to investigate the potential of PV and optimize the use of available solar energy resources. The findings will provide insight into the viability of solar power in regions with moderate temperatures as well. The findings are expected to offer valuable insights into the economic and storage implications of harnessing solar energy in such regions.

In this paper, the Ludington PSH is considered the upper reservoir of the hydro facility due to its unique characteristics as one of the largest hydroelectric plants in the United States. The plant works by pumping water from the lake to a reservoir approximately 110m above the shoreline. The reservoir is about two miles long, 33.528m deep, and can hold 94.63 million $m^3$ of water [9]. When water is released back downhill, the reservoir level can drop by up to 21.3m over several hours. As the water flows down, six turbines in the powerhouse generate up to 2,322 MW of electricity, which is enough to power around 1.6 million people for a certain period of time and helps stabilize the upper Midwest's power grid [9].

### III. MATHEMATICAL ANALYSIS AND DESIGN STRATEGY

In this section, a brief mathematical modeling overview will be presented, detailing how data is processed within the PV system to calculate key parameters, such as hourly solar irradiances. The hourly horizontal solar-measured real data from the proposed Ludington PSH, as presented previously in Section III, will be employed in modeling the output power for the PV panel. Note that the modeling of solar angles, mPV, and bPV is explained in detail in [10], providing further insight into the precision and methodology involved in the system's design.

In this study, mPV and bPV modules were selected for comparative analysis with bPV options usually paired with monocrystalline technology. The mPV and bPV modules used in this study were sourced from Canadian Solar's module lineup. The mPV panel has a rated power of 420W with an efficiency of 18.8%, while the bPV panel achieves a rated power of 462W. Notably, the efficiency of the bPV module increases by around 10%, reaching 20.68%. Further details on these modules can be found in the datasheet presented in [10].

#### A. Solar Data Processing and Output Irradiances for mPV Module

The data must first be converted into the appropriate form. Therefore, $GHI(t)$ for the *mPV* panel could be found using (1), the hourly $GHI$ is mathematically derived from $DHI(t)$, $DHI(t)$, as mentioned in section III, and the zenith angle $\theta_z(t)$. With the $GHI$ data computed, the global tilted irradiance for mPV ($I_{Gtm}$) is then ready to be incorporated into the PV module modeling as will be discussed in Section V. $I_{Gtm}$ is illustrated in Fig. 5 and computed at each time step using (2) [11].

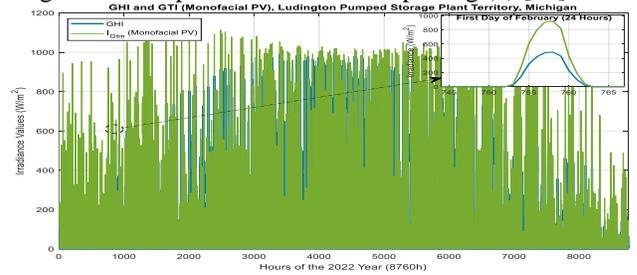

Fig. 5. Annual GHI and GTI for mPV system at Ludington pumped storage area

$$GHI_m(t) = DHI(t) + DNI(t) \times \cos(\theta_z(t)) \quad (1)$$

$$I_{Gtm}(t) = GHI(t) \times \frac{\sin(\alpha_s(t) + \beta_s)}{\sin(\alpha_s(t))} \quad (2)$$

Here, $\alpha_s$ is the solar elevation angle, $\varphi$ is the latitude and the declination angle (δ). For the Ludington PSH territory, $\beta_s$ is set to 36.5° for the stationary PV panels [8], and the latitude φ is 43.893648°, based on the coordinates of the selected site. Note that, the detailed derivation is can be found in [11].

*B. Solar Data Processing, and analysis for bPV Module*

The bPV module was selected as a new PV technology to assess its impact on the PSH system compared to the conventional mPV technology. The bPV solar panels are designed to capture irradiance from both the front and rear surfaces by factoring in the global beam ($I_{G_{B_{Beam}}}$), diffused ($I_{G_{B_{Diff}}}$), and reflected irradiances ($I_{G_{B_{Ref}}}$), respectively. Using the input solar data provided by the manufacturer and the specifications from the selected PV panel datasheet, these values were used to calculate the total tilted irradiance for bPV panel ($I_{G_B}$), using (3). The solar panel irradiances for the tilted configuration are depicted in Fig. 6. Note that, the detailed mathematical derivation of this can be found in the [10].

$$I_{G_B} = I_{G_{B_{Beam}}} + I_{G_{B_{Diff}}} + I_{G_{B_{Ref}}} \tag{3}$$

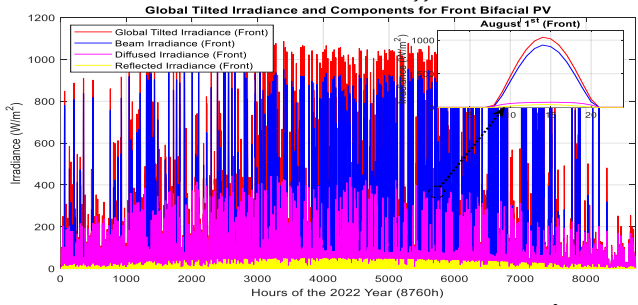

Fig. 6. Hourly computations of front-tilted bPV irradiance in $W/m^2$.

The results show that bPV performs better than mPV because of its ability to capture irradiance from both the front and rear sides. For mPV, the GHI has a maximum of around 1 kW/m² with a mean of 153.10 W/m², while the maximum tilted irradiance ($I_{Gtm}$) reaches 1,113.3 $W/m^2$. In contrast, bPV shows enhanced performance with front irradiance peaking at 1,129.3 W/m² and rear irradiance contributing up to 350.66 $W/m^2$. This results in a total irradiance ($I_{Gt_B}$) of 1,330.2 W/m², significantly higher than mPV's maximum. Fig. 7 illustrate the rear-side tilted irradiance values which are lower than those on the front. It's worth mentioning that these values are calculated for each time step, taking into account the sun's changing altitude at this location over the 8,760 hours of 2022.

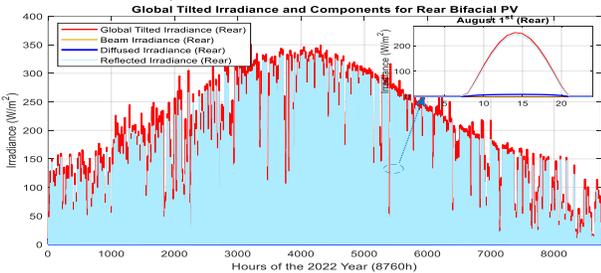

Fig. 7. Hourly computations of rear-tilted bPV irradiance in $W/m^2$.

Furthermore, as illustrated in Table II and Fig. 8, the average $I_{Gt_B}$ for bPV (217.91 $W/m^2$) surpasses that of mPV (193.95 $W/m^2$) by approximately 12.37%, with the peak irradiance for bPV being around 19.5% higher than mPV. This leads to a significant decrease in the number of PV panels needed for the entire solar system, especially in large-scale megawatt projects.

TABLE II. Minimum, average, and maximum computed irradiance values (w/m²) for mPV and bPV technologies at the selected location in 2022.

| Type of PV Technology | | | Summary Statistics in ($W/m^2$) | | |
|---|---|---|---|---|---|
| | | | Min | Mean | Max |
| Monofacial (mPV) | | GHI | 0 | 153.0991 | 1,002 |
| | | $I_{Gtm}$ | 0 | 193.9523 | 1,113.3 |
| Bifacial (bPV) | Front | $I_{G_{B_{Beam}}}$ | 0 | 115.1686 | 1,002.4 |
| | | $I_{G_{B_{Ref}}}$ | 0 | 7.5073 | 49.1339 |
| | | $I_{G_{B_{Diff}}}$ | 0 | 53.0492 | 441.9449 |
| | | $I_G$ Front | 0 | 175.7252 | 1,129.3 |
| | Rear | $I_{G_{B_{Beam}}}$ | 0 | 0 | 0 |
| | | $I_{G_{B_{Ref}}}$ | 0 | 54.5004 | 337.5899 |
| | | $I_{G_{B_{Diff}}}$ | 0 | 5.7683 | 48.0551 |
| | | $I_G$ Rear | 0 | 60.2688 | 350.6634 |
| Total (Front+Back) $I_{Gt_B}$ | | | 0 | 217.9133 | 1,330.2 |

The data demonstrates that bPV's capacity to absorb both direct and reflected light leads to a significant boost in energy production. Despite Michigan's colder climate, these findings highlight bPV's potential as a more efficient option, particularly in areas with high reflected irradiance. While these results are based on a single PV array, they raise important questions about the implications for utility-scale, grid-connected systems. The comparative analysis of mPV and bPV modules highlights their unique strengths and limitations. bPV demonstrates better energy production, particularly in high-albedo environments. These insights aim to guide stakeholders in selecting the most suitable technology based on project-specific goals, constraints, and environmental factors.

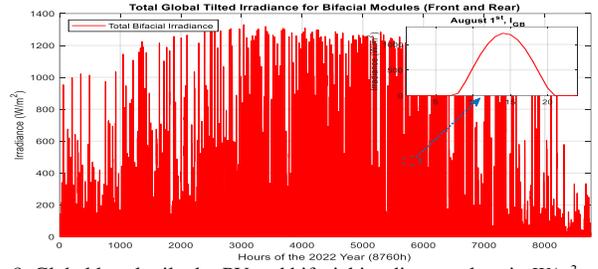

Fig. 8. Global hourly tilted mPV and bifacial irradiance values in $W/m^2$.

## IV. ENERGY MANAGEMENT AND SYSTEM'S OPERATIONAL FRAMWORK: OPTMIZATION OBJECTIVE FUNCTION

Several research studies have explored similar operational schemes for such systems. In [12], the power generation strategy for a PV system integrated with grid-connected PSH is outlined. This study focuses on optimizing the system at each time interval using the Multi-Objective Evolutionary Algorithm based on Decomposition (MOEA/D). The comprehensive mathematical approach is thoroughly discussed in [11], covering power and energy management, including the equality equations, inequality constraints, and the methodology employed. This same approach will also be taken in this paper.

In this paper, the objective function (OF) is designed to address three key aspects of the problem: cost minimization, renewable energy usage maximization, and storage utilization optimization. These three objectives will work together to comprehensively solve the problem at hand. The renewable storage usage factor (RSF) evaluates how much energy from the upper Ludington PSH of the hydropower facility meets the total demand, as shown in equation (4) [12]. Here, *E*<sub>storage</sub> represents the energy delivered by the PSH to the load, while *E*<sub>System</sub> includes the total energy output of the system, including PSH contributions. The LCOE computed as in (5) is the 2<sup>nd</sup> objective function; cost OF is used to estimate the average cost of generating electricity over the system's lifetime. Minimizing the LCOE is a primary goal to ensure economic feasibility. It incorporates initial capital costs (e.g., equipment, and installation), operational and maintenance expenses (routine servicing and repairs), and replacement costs (such as inverters), alongside the system's total energy output over its lifespan [13]. By accounting for these factors, the LCOE provides a comprehensive measure of the cost-effectiveness of the system. The third objective aims to maximize the Index of Reliability (IR), which evaluates the system's ability to meet load demands without interruptions. IR ensures a consistent and reliable energy supply and is calculated using equation (6).

$$SOF: Max\left(RSF = \sum_{0}^{t}\frac{E_{Storage}}{E_{System}}\right) \quad (4)$$

$$COF: Min\left(LCOE = \frac{ACS}{E_s}\right) \quad (5)$$

$$ROF: Max\left(IR = 1 - \frac{\sum_{t=1}^{8760}[P_L(t)-(P_{PV_{inv}}(t)+P_{WT}(t)+P_{ydro_{turbine}}+P_{gp}(t))]}{\sum_{t=1}^{8760}P_L(t)}\right) \quad (6)$$

Where $P_{PV_{inv}}$, $P_{WT}$, $P_{ydro_{turbine}}$, and $P_{gp}$ are the inverted solar, wind turbine, hydro discharged and grid purchased power values at each time step. Together, these objectives, minimizing system costs (through LCOE), maximizing RSF, and maximizing reliability (IR), will allow for the best estimation of the impact of mPV and bPV solar systems on the Ludington PSH system. Equation (7) presents the three OFs alongside the decision variables, namely the number of solar PV panels ($N_{PV}$) and hydro turbines ($N_{Ht}$), subjected to equality and inequality constraints similar to those outlined in [12]. This approach ensures optimal storage usage, the lowest possible system costs, and maximum reliability, which are critical to the success of large-scale renewable energy projects.

$$(SOF \& COF \& ROF) = f(N_{PV}, N_{Ht}) \quad (7)$$

The MOEA/D was selected for this study due to its strong ability to handle the complexities of our problem, particularly in managing multiple, often conflicting objectives [13]. Aiming to optimize three objectives including minimizing costs, and maximizing the use of RESs, and improving storage efficiency, all while ensuring system reliability. MOEA/D can handle these three objectives effectively, dividing the multi-objective problem into scalar subproblems for each one objective, and optimizing each subproblem individually while ensuring diversity across Pareto-optimal solutions. This method is particularly well-suited for managing the complexity of integrating renewable energy sources, storage, and grid reliability due to its flexibility in addressing nonlinearity and conflicting objectives. MOEA/D ensures the simultaneous optimization of cost minimization, renewable energy maximization, and storage efficiency [14]. Additionally, it enables efficient management of large search spaces by dynamically exploring various combinations of the decision variables ($N_{PV}$ and $N_{Ht}$) to identify the best configurations.

## V. RESULTS AND DISCUSSION

After processing the data, this section applies the mathematical modeling and subsequent energy management strategies for the Ludington PSH area to determine the optimal configuration for the mPV and bPV systems. Table III displays the optimal sizing configuration for both solar grid-connected systems.

Although bPV panels are more expensive, their reduced numbers and installation requirement impacts the overall system cost and storage. Table III show that bPV technology is more efficient and cost-effective than mPV. bPV requires fewer PV panels (3784 vs. 4108), has a higher IR (99.037% vs. 97.270%), and achieves a lower LCOE ($0.0578/kWh vs. $0.0672/kWh). Additionally, bPV has a slightly higher RSF (38.071% vs. 37.682%). This suggests that the bPV system operates more efficiently in meeting demand, with lower costs and higher storage contributions. Fig. 9 illustrate the MOEA/D triple OF for bPV case.

TABLE III. Optimum values of the OF's and the decision variable for both cases.

| Cases | $N_{PV}$ | $N_{Ht}$ | IR (%) | LCOE ($/kWh) | RSF (%) |
|---|---|---|---|---|---|
| mPV | 4108 | 5 | 97.270 | 0.0672 | 37.682 |
| bPV | 3784 | 5 | 99.037 | 0.0578 | 39.071 |

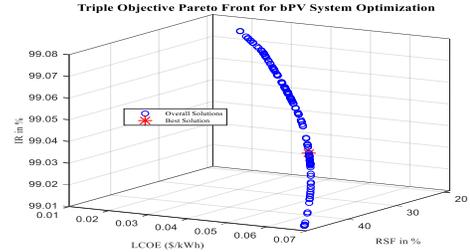

Fig. 9. Optimization of bPV System: Triple objective Pareto front analysis.

Table IV highlights key differences between the mPV (more solar panels with higher LCOE) and bPV (fewer solar panels and lower LCOE) configurations in terms of energy values. The bPV case exhibits a higher inverted energy at 3.3428GWh, indicating enhanced PV utilization. The hydropower turbine energy output is also greater at 2.2643GWh, contributing positively to the overall efficiency. Notably, the pumping energy required for bPV is higher at 0.9339GWh, which suggests a greater reliance on pumped storage to optimize energy flow and balance supply with demand. Additionally, the bPV system pumps 10.38% more water annually to the upper Ludington PSH than the mPV system, enhancing overall energy generation. This enables the system to store more water and generate additional energy during peak demand, as seen in the higher $E_{hydro_{turbine}}$ values. In summary, the bPV system, with fewer solar panels, delivers better energy inversion and water pumping performance, making it a more efficient and cost-effective configuration.

TABLE IV. Energy values and water pumping comparison between mPV and bPV configurations.

| Energies (GWh) | mPV | bPV |
|---|---|---|
| $E_{inv}$ | 3.2303892 | 3.342776 |
| $E_{hydro_{turbine}}$ | 2.203793 | 2.264277 |
| $E_{pump}$ | 0.9129272 | 1.28144735 |
| $E_{gpurch}$ | 0.363040995 | 0.36718031 |
| $E_L$ | 4.1052819 | 4.1052819 |
| $E_{gsold}$ | 0.771495396 | 0.66754789 |
| $E_{deficit}$ | 0.008081316 | 0.00824943 |
| Total pumped water in million m³/year | 4.8382 | 5.3407 |

Fig. 10 displays the inverted PV output power for MW's bPV and mPV systems. The bPV system is observed to reach a maximum inverted power value of around 2.33MW, compared to 2.12MW for the mPV system as illustrated in Table V. Additionally, the average generated power throughout the year is approximately 0.38MW for the bPV system and 0.37MW for the mPV system.

Despite a decrease of 8.21% in the number of solar panels for the bPV system compared to the mPV system, the bPV still delivers higher mean and maximum output power. This highlights the efficiency of bPV technology, which benefits from the ability to absorb solar irradiance from both sides of the panels. Consequently, the bPV system produces more energy using fewer solar panels, with enhanced its water storage capabilities. The results in Fig. 11 and Table V, indicate that the bPV system demonstrates a better impact on energy storage compared to the mPV system.

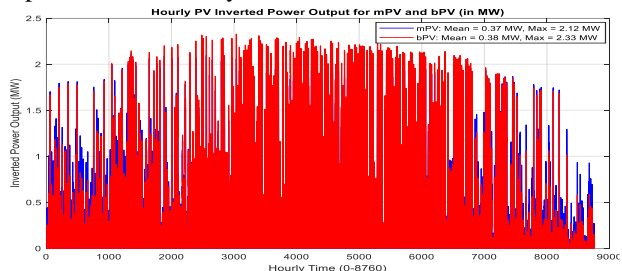

Fig. 10. Hourly performance of mPV and bPV inverted power (MW).

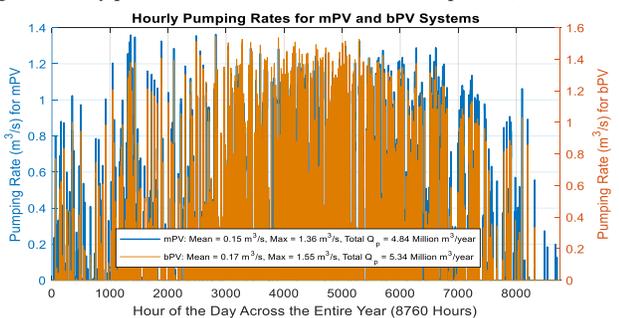

Fig. 11. Comparison of hourly pumping rates for mPV and bPV systems.

TABLE V. Minimum, average, and maximum values of inverted solar power and pumping rates for mPV and bPV systems.

| Type of PV Technology | | Value | | |
|---|---|---|---|---|
| | | Min | Mean | Max |
| $P_{inv}$ (MW) | mPV | 0 | 0.37 | 2.12 |
| | bPV | 0 | 0.38 | 2.33 |
| Pumping$_{Rates}$ (m$^3$/s) | mPV | 0 | 0.15 | 1.36 |
| | bPV | 0 | 0.17 | 1.55 |

With a higher mean pumping rate, the bPV system can pump a total of 5.34 million m$^3$ of water per year. This is approximately 10% more than the mPV system. The increased pumping capacity of the bPV system reflects its ability to store more energy, making it more efficient in utilizing surplus power for energy storage in the pumped hydro system.

TABLE VI. Sensitivity Analysis of Key Parameters and System Performance for mPV and bPV Technologies

| Varying parameters | Cases | $N_{PV}$ | $N_{Ht}$ | IR (%) | LCOE ($/kWh) | RSF (%) |
|---|---|---|---|---|---|---|
| GHI (+10%) | mPV | 3976 | 4 | 97.41 | 0.051 | 39.42 |
| | bPV | 3512 | 4 | 99.45 | 0.048 | 41.56 |
| GHI (-10%) | mPV | 4468 | 6 | 96.23 | 0.083 | 35.81 |
| | bPV | 3941 | 6 | 98.73 | 0.069 | 36.16 |
| $T_{amb}$ (+10%) | mPV | 4263 | 5 | 97.64 | 0.0612 | 37.682 |
| | bPV | 3523 | 5 | 99.31 | 0.0503 | 40.27 |
| $T_{amb}$ (-10%) | mPV | 4326 | 5 | 96.98 | 0.074 | 36.27 |
| | bPV | 3853 | 5 | 98.94 | 0.0607 | 38.42 |

Table VI presents the results of sensitivity analysis on key parameters, including solar irradiance and ambient temperature, to evaluate their impact on the main system performance metrics. An increase in GHI by 10% reduces $N_{PV}$ and improves LCOE, with bPV achieving $0.048/kWh compared to $0.051/kWh for mPV. However, higher $T_{amb}$ slightly improves IR and slightly lowers LCOE. bPV consistently performs better than mPV scenarios, including lower LCOE and higher RSF, under all conditions. These results underscore bPV's robustness and efficiency, making it a better choice over mPV in varying conditions.

## VI. CONCLUSIONS

This study compares traditional mPV and bPV solar systems integrated with PSH, focusing on their performance in grid-scale hybrid renewable systems in Ludington, Michigan. Using real-world data, the research highlights the advantages of bPV, which generates more energy with fewer panels and a lower LCOE over its lifespan. By optimizing factors including cost, renewable energy usage, and storage efficiency, the study employs MOEA/D to address these objectives. The bPV system's ability to pump more water increases flexibility and energy availability, enabling higher output during turbine operations. This improved pumping capacity enhances system performance and efficiency. In summary, the 10.38% increase in water pumping for the bPV system leads to higher energy production and better operational efficiency than mPV.

## ACKNOWLEDGMENT

This work was supported in part by the National Science Foundation of USA under Grant ECCS-2146615 and partially supported by the Department of Energy, Solar Energy Technologies Office Renewables Advancing Community Energy Resilience (RACER) program under Award Number DE-EE0010413. Any opinions, findings, conclusions, or recommendations expressed in this material are those of the authors and do not necessarily reflect the views of the Department of Energy.